\documentclass[10pt,a4paper]{article}

\usepackage[utf8]{inputenc}
\usepackage[T1]{fontenc}
\usepackage{microtype}
\usepackage[sc]{mathpazo}
\usepackage[dvipsnames]{xcolor}
\usepackage[hyphens]{url}
\usepackage{hyperref}
\hypersetup{
    colorlinks=true,
    allcolors=NavyBlue
}

\usepackage{graphicx}

\usepackage[bottom, hang]{footmisc}
\usepackage[
    top=2.5cm,
    bottom=2.5cm,
    left=2cm,
    right=2cm,
    footskip=1cm,
    headsep=0.75cm,
    columnsep=20pt,
]{geometry}

\usepackage{fancyhdr}
\patchcmd{\maketitle}{plain}{empty}{}{}

\usepackage[english]{babel}
\usepackage{apacite}
\usepackage[authoryear,round]{natbib}

\usepackage{titling}
\pretitle{\begin{center}\huge\bfseries}
\posttitle{\end{center}}
\patchcmd{\maketitle}{plain}{empty}{}{}

\usepackage{booktabs}
\usepackage{rotating}
\usepackage{makecell}

\usepackage{caption}
\usepackage{enumitem}
\setlist{noitemsep}
\usepackage{amsmath,amssymb}

\title{Distorted Perspectives of LLM-Simulated Preferences: Can AI Mislead Design?}

\author{Eduard Kuric\textsuperscript{1,2}\thanks{Corresponding author: \href{mailto:eduard.kuric@stuba.sk}{eduard.kuric@stuba.sk}\\ORCID(s): 0000-0002-7371-5512 (E. Kuric), 0000-0002-4111-1052 (P. Demcak), 0000-0001-9030-7337 (M. Krajcovic)\\}, Peter Demcak\textsuperscript{2} and Matus Krajcovic\textsuperscript{1,2}
}

\date{
\footnotesize\textsuperscript{1}
Faculty of Informatics and Information Technologies, Slovak University of Technology, Ilkovicova 2, Bratislava, 84216, Slovakia\\
\textsuperscript{2}
UXtweak Research, UXtweak j.s.a., Cajakova 18, Bratislava, 81105, Slovakia\\
}

\begin{document}

\maketitle

\begin{center}
\normalfont\bfseries\vspace{0.5\baselineskip} \abstractname
\end{center}
\begin{quote}
\normalfont\small
\textbf{Purpose.} Designers of digital solutions increasingly consult Large Language Models (LLMs) for their work. However, it remains unclear how this may affect the user experiences they produce and there are no established practices. We investigate how design preferences expressed by LLM-driven simulation methods align with those of real users.\\
\textbf{Design/methodology/approach.} We present a study that aggregates real-world data and design stimuli from twenty-nine preference tests conducted in practice by users of the UXtweak online research platform (n = 2073). We perform holistic multimodal simulations where we manipulate LLM variables (model reasoning, sampling, persona type, and specificity) and assess their effects on algorithmic fidelity.\\
\textbf{Findings.} Our results unveil significant and systematic discrepancies between peoples’ real design preferences and LLM simulations that are consistent across manipulations. Synthetic justifications lack genuine depth, nuance and reasoning, which they substitute by patterns like  focus on generic properties, specific elements, elaboration and overpraising.\\
\textbf{Originality/value.} The unique attention directed by this research toward preferences within visual design stimuli highlights misrepresentation of perception and meaning by LLMs in a context that is intuitive yet critical for design teams. The external and ecological validity of our findings is high, given their replication across a multitude of real-world studies.  
\end{quote}

\begin{quote}
\end{quote}

\section{Introduction}\label{sec:introduction}

The rapid proliferation of generative artificial intelligence (GenAI) models — LLMs most prominently — has had deep social impact. In spite of broad industrial pressures to adopt GenAI that can result in what \citet{niedehoffer2025} termed as “workslop”, a survey by \citet{takaffoli2024} suggests a lack of clear guidelines for user experience (UX) practitioners responsible for designing digital solutions. Online research platforms such as UXtweak\footnote{UXtweak user experience research platform: \url{https://www.uxtweak.com/}}, as well as researchers and practitioners have started exploring their potential for UX and interface design \citep{takaffoli2024}. Experimental applications include creation of personas \citep{schuller2024}, inference of user needs and goals \citep{zhu2025}, heuristic evaluation \citep{duan2024}, prototyping tool plugins for designing mockups \citep{petridis2023} and ideation \citep{kim2022}. Use of GenAI can make creative outputs appear more desirable \citep{zhu2026}. The consensus of belief among UX practitioners is that AI can improve efficiency, although it is tempered by caution over the express need to verify GenAI outputs, as well as concerns about overdependence and reduction of critical thinking \citep{takaffoli2024}. The most common application of GenAI by designers is in text-oriented scenarios, where LLMs are used for brainstorming and as a source of second-opinion-style feedback \citep{takaffoli2024}.

LLMs playing the advisory role for UX designers is where we identify the issue central to our research. While designers of digital solutions may separate themselves from their own proposals when making decisions, they evaluate concepts with a confirmation bias \citep{nikander2014}. If relied on, LLMs could magnify biases and legitimize flawed conclusions by providing external validation for ideas that might be at odds with preferences derived through genuine research, critical thinking, and empathy \citep{klingbeil2024}. Preferences can be defined as dispositions based on beliefs (e.g., expectations, values, needs, aesthetic sentiments), operating on the threshold between cognition and behavior \citep{guala2019}. They are granular and uniquely personal due to the diversity of life experiences \citep{yoon2024}. In UX design, user preferences can affect the quality and success of digital products and services by shaping the willingness of people to use them, the manner of using them, their subjective experience, and the resulting satisfaction. Design preferences may be reflected in use efficiency \citep{baughan2020}. Therefore, practices involving sloppy design advice \citep{niedehoffer2025} that misrepresent human preferences could pose serious harm to organizations. It might compromise the usefulness and value of their digital solutions while inflating their development costs, especially when used in early stages.

Our study addresses the research gap concerning the lack of understanding of how LLMs mirror human design preferences. Simulation literature refers to this type of accuracy using the term algorithmic fidelity \citep{xie2024}. Although LLMs are fine-tuned through reinforcement learning from human feedback (RLHF) to make the output itself in line with user preference (e.g., more structured or factual) \citep{sparrenberg2024, payne2026}, this does not inherently translate into an ability to reproduce contextually aware preferences on their subject. Models may present distortions such as agree bias \citep{dorner2023}, prosocial preference \citep{xie2024}, and minimal variability \citep{park2024c}. Previous studies demonstrated inconsistencies with broader social or consumer surveys \citep{durmus2024, brand2024}. The focus of our study lies in multimodal evaluation of design preferences where LLMs are provided with concrete sets of visual stimuli as context with high degree of detail. When granted equal information to human samples and steered to simulate them based on descriptive information, how well do LLMs extract comparable implications about preferences for design teams?

To contribute insights with generalizable implications for theory, real-world UX practice, and organizational policies, we first acquired a diverse set of twenty-nine (29) preference testing studies — each by a different organization, amounting to 2073 participants in total. Then, we used this dataset as a source of ground truth for LLM simulations involving six (6) different variants to comprehensively assess the capacity of different methods to improve simulated preferences. Alarmingly, our findings demonstrate a systematic failure of LLMs to simulate human design preferences that applies universally across different studies. Furthermore, advanced models with chain-of-thought reasoning, persona modeling, temperature and nucleus sampling have minimal impact on the underlying issues. Our analysis advances the understanding of deviations between human and LLM-simulated design choices, as well as patterns that differentiate generated justifications from human experiential narratives.

\section{Background and literature review}\label{sec:background}

\subsection{Evaluation of design preferences}

To echo self-determination theory, the consumption of design by users is not passive \citep{stock2014}. Human preferences — toward software, travel, food or any other choice domain — are driven by intrinsic and extrinsic motives in pursuit of both utilitarian and hedonic value \citep{stock2014}. Relevant factors can include attitudes and beliefs \citep{russo2019}, perceived usability and aesthetics \citep{lee2010}.

Preference can be evaluated through choice behavior \citep{pfefferle2025}. In user experience research, design preferences are typically gauged through preference testing. This is an established and popular technique where participants are asked to select their preferred design among two or multiple options \citep{tomlin2018}. Ensuing follow-up questions ascertain their justification (why?) alongside other information. This method is a simple and quick source of both qualitative and quantitative feedback. The prevalence of research platforms like UXtweak has made online preference testing easy and widely accessible.

\subsection{Representation of preferences in LLMs}

With advancements in AI models, researchers have probed LLMs’ ability to reflect human preferences based on the information from the internet (such as reviews) encoded in their semantic embeddings \citep{qu2024, durmus2024, santurkar2024}. Unlike problematic earlier models, modern LLMs undergo reinforcement learning from human feedback (RLHF) to make them more helpful, harmless and honest \citep{li2025b, dorner2023, santurkar2024}. The consequence is more bias in their outputs, including preferences. The focus of many studies was therefore to examine whether intervening with prompts and fine-tuning can steer simulations to become more accurate.

Latent preferences — manifestations of values and judgements toward economic, political, moral or environmental issues — were the most common type studied in LLM simulations \citep{durmus2024, qu2024, santurkar2024, park2024c}. Despite some alignment with dominant preferences from USA-centered public opinion surveys, which can be attributed to high representation in training data, different regions and specific demographics (e.g., age, education) demonstrated significant distortions \citep{qu2024}. Using another country’s native language does not increase fidelity while invoking its name can induce harmful stereotypes \citep{durmus2024}. Even after steering towards different ends of the political spectrum, LLMs show liberal tendencies but with a right-leaning moral foundation \citep{santurkar2024, park2024c}. Option order can be a decisive factor for some preferences, while others received deterministic answers regardless of what group was being simulated, comparably to humans solving 2 + 2 \citep{park2024c}. 

The failure of LLM simulations to adopt latent human characteristics and biases and draw appropriate implications while answering questions \citep{giorgi2024} suggests that simulation of design preferences is likely to struggle with accounting for top-down factors involved in their mental formation. Although image processing may provide accurate descriptions of visual designs, failure to form human-like associations to translate into preferences could also hinder transformation into preferences from the bottom-up \citep{imschloss2025}.

Previous studies of LLM approximating consumer preferences towards products or entertainment media showed mixed results \citep{brand2024, yoon2024}. Willingness to pay for product features mapped to the general audience but not across specific demographics \citep{brand2024}. Model fine-tuning helped achieve better alignment for features within the same product, but had a negative impact between products \citep{brand2024}. While simulating preferences for movies, LLMs presented as unrealistically flexible in accepting recommendations \citep{yoon2024}.

\section{Research model}\label{sec:research-model}

\begin{figure}[ht]
\centering
\includegraphics[width=0.8\textwidth]{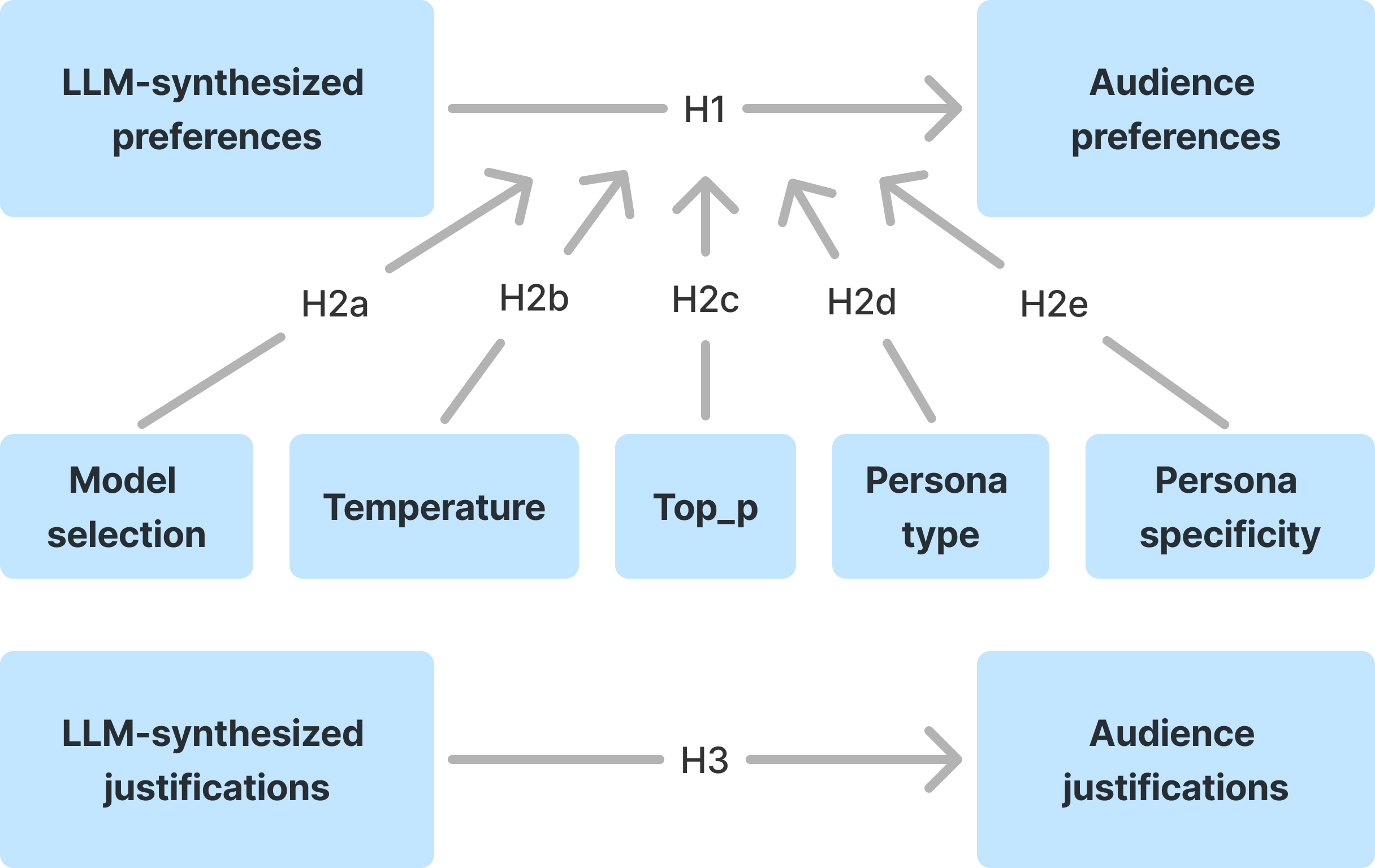}
\caption{Research model.}\label{fig:research-model}
\end{figure}

The risk of LLM misalignment with human design preferences presents the focal object of our research. Accordingly, our research model maps the relationships between synthetic and genuine audience as shown in \autoref{fig:research-model}. Our hypotheses posit that divergences will emerge in quantitative voting results (\hyperref[hyp:h1]{H1}) and their qualitative reasoning justifications (\hyperref[hyp:h3]{H3}). Parameters of solutions that have been suggested as model improvements are assessed as moderating variables (\hyperref[hyp:h2a]{H2a}-\hyperref[hyp:h2e]{H2e}).

\phantomsection\label{hyp:h1}
Recent studies have indicated that despite LLMs successfully generating text that may appear real \citep{lazik2025}, they fall short of matching human subjectivity, experiences, diversity and biases \citep{giorgi2024}. As stochastic models that generate likely answers, they were found to rely on stereotypes \citep{lazik2025}. We expect that preferences — born from the interplay between latent factors of design and subjectivity — will elude reliable prediction.\\

\noindent
\emph{H1.}\\ 
LLM-synthesized design preferences do not predict audience preferences.\\

\phantomsection\label{hyp:h2a}
\phantomsection\label{hyp:h2b}
\phantomsection\label{hyp:h2c}
More advanced training and tuning may not address intrinsic issues manifested heterogenously across the diversity of available models \citep{brand2024, gao2025}. So-called Large Reasoning Models create a facsimile that collapses in deeper problems \citep{shojaee2025}. To assess their prospect, we thus focus on two models: GPT 5.2 — touted as the best reasoning model \citep{openai2025a} — and GPT 4.1, an efficient non-reasoning alternative \citep{openai2025b}. Various versions of GPT are the most investigated LLM simulation and comparable to alternatives such as Claude or LLaMa \citep{hamalainen2023, argyle2023}. Temperature and top\_p (nucleus) are sampling parameters that control entropy and thus the variability of the outputs. Since they may not result in genuine semantic and thematic diversity \citep{petrov2024, ma2025b}, we postulate that their impact on synthetic preferences will be minimal.\\

\noindent
\emph{H2a.}\\ 
Model selection does not affect the similarity between LLM-synthesized and audience design preferences.\\

\noindent
\emph{H2b.}\\ 
Temperature does not affect the similarity between LLM-synthesized and audience design preferences.\\

\noindent
\emph{H2c.}\\ 
Top-p does not affect the similarity between LLM-synthesized and audience design preferences.\\

\phantomsection\label{hyp:h2d}
\phantomsection\label{hyp:h2e}
Works simulating participants have imposed various persona representations to prime models toward better alignment with audiences \citep{gerosa2024}. Personas can represent individuals (single-persona) \citep{hamalainen2023}, but also groups (multi-persona) \citep{zhang2024} or populations (mega-persona) to increase diversity \citep{argyle2023}. Enriching personas with information about demographics (e.g., age, gender education) \citep{petrov2024, brand2024}, personality \citep{sparrenberg2024, petrov2024, dorner2023, li2025b} or other factors of individual experiences \citep{petrov2024, li2025b} has demonstrated limited effects on distortions. For this reason, we do not expect personas to significantly moderate the fidelity of design preferences toward realism.\\

\noindent
\emph{H2d.}\\ 
Mega-persona LLM-simulated preferences are more similar to audience preferences than single-persona preferences, but the similarity is partial.\\

\noindent
\emph{H2e.}\\ 
Specific descriptive personas do not affect the similarity between LLM-synthesized and audience design preferences.\\

\phantomsection\label{hyp:h3}
For deeper understanding of synthetic design preferences, we investigate LLM answers to follow-up questions as a source of insight about preference reasoning. Given that LLMs generate details of their explanations in superposition over existing context instead of having latent thoughts that they could report \citep{shanahan2023}, the term “justification” may more aptly represent the comparison between self-reported reasoning and LLM rationalization. Since prior research reported stereotypicality and inconsistency in lieu of human nuance and deep experiential subjective diversity \citep{venkit2025}, we hypothesize: \\

\noindent
\emph{H3.} \\ 
LLM-synthetized justifications of design preferences do not predict audience justifications.\\

\section{Methodology}

\subsection{Participants}

To maximize realism and generalizability, we collected a large dataset that collates twenty-nine (29) diverse preference testing studies. The studies were conducted in UXtweak, an online platform used by organizations worldwide to support various UX research methods, including evaluation of designed prototypes \citep{kuric2025hotspots} or information architectures \citep{kuric2025tt}. Practitioners can use the Preference Test tool to create custom studies and gauge their audience’s preferred design selections, recruit participants and collect feedback with supporting questions. UXtweak obtained the consent of randomly selected users to access authentic studies from their real practice. The results comprise anonymized ground truth data from 2073 participants, 78 tasks and 190 follow-up question justifications (147 open- and 43 close-ended). Original domains encompass travel \& transportation, health \& wellness, media \& entertainment, business \& finance, living, education, and infrastructure. The evaluated designs include user interfaces and their components (mainly from web or mobile apps), layouts, notifications and navigation components, as well as other visual elements, like illustrations and copywriting.

As justifications for hypothesis \hyperref[hyp:h3]{H3} of this study, we consider answers that are both explicit and implicit. Questions asking for explicit explanations of preferences (e.g., “Why did you choose this design?”) were found in 36\% of tasks. Other questions regarding preferences included implicit justifications that provide broader context to support the decision (e.g., ”What did the option you selected make you think?”).

\subsection{Procedure}

\begin{figure}[ht]
\centering
\includegraphics[width=\textwidth]{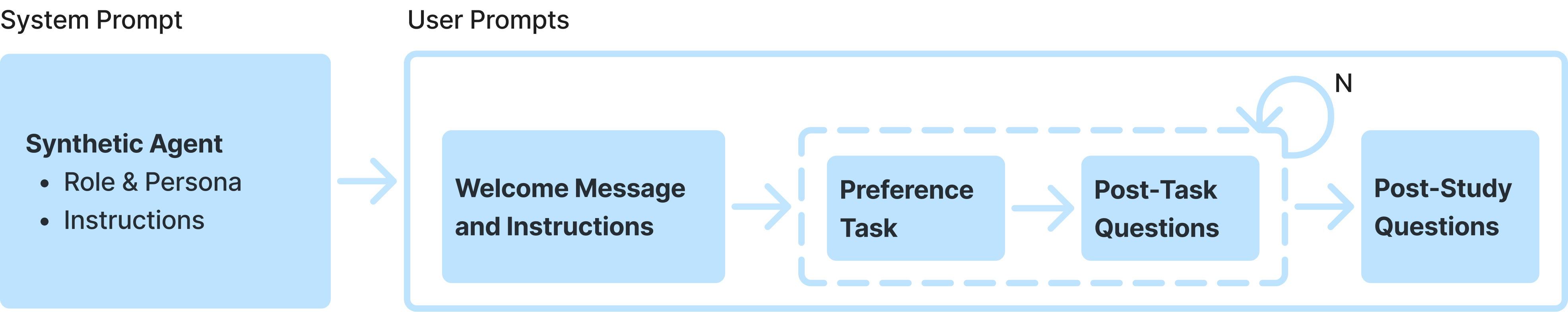}
\caption{Preference test LLM simulation procedure.}\label{fig:procedure}
\end{figure}

LLM-synthesized preferences were generated by simulating the participants’ standard user flow with contents of acquired preference tests. First, the multi-stage procedure shown in \autoref{fig:procedure} established the LLM’s role and instructions to simulate participants in the system prompt. Then, user prompts guided the LLM through steps of the activity, with general directives, user-written messages to mirror the experience of human participants and output format specification. Since LLMs are stochastic models, evaluation across multiple runs can more accurately capture their behavior \citep{ma2024}. However, we found that upon synthesizing all preference test studies three times using the baseline configuration (GPT 4.1, temp=1, top\_p=1, specific mega-persona), absolute values exhibited minimal variations between iterations. Statistical testing confirmed no significant differences across the runs, furthermore differences were limited even between different settings as we discuss below. Therefore, to prevent redundancy, single runs were used as reliable and clear representations of model behavior. This approach saves costs associated with repetitive simulations of already numerous studies and configurations.

Our ensemble of hypotheses demanded that the simulations be performed iteratively with different settings. We used GPT 4.1 as the baseline model to assess LLM-generated design preferences and their justifications, with a parameter configuration intended to improve its algorithmic fidelity. Mega-personas and recommended values of temperature and top\_p = 1 were used to stimulate output variability \citep{gerosa2024, ma2025b}. Temperatures greater than 1 were rejected, as their enhancement of diversity can be limited and come at the cost of coherence \citep{ma2025b}.

Information about sample size and personas was drawn from real study data (panel recruitment criteria, screening questions, pre-study questionnaire answers). For studies not explicitly aimed at specific demographics and personality profiles, we augmented the personas with empirical distributions of gender, age\footnote{Age distribution: \url{https://www.statista.com/statistics/272365/age-distribution-of-internet-users-worldwide}}, education\footnote{Education attainment: \url{https://www.oecd.org/en/topics/education-attainment.html}} and Big-5 personality types in general user populations \citep{fleeson2010}. Sensitivity analysis in hypotheses \hyperref[hyp:h2a]{H2a}-\hyperref[hyp:h2e]{H2e} was supported by study simulations with the change of a single variable: either model to GPT 5.2, temperature to 0.2, top\_p to 0.2, persona type to single-persona, or persona specificity removed). For example, single-persona simulation contains persona specifications that characterize individuals rather than the population, with the baseline model and parameters.

For mega-persona simulations, personas were aggregated to summarize the audience’s distribution of demographics, personality traits and other descriptive information available from the study (e.g., experiences, knowledge, practices, habits). Single-persona conversations as simulacra of participants were steered through personas that conveyed their individual details. User prompts were zero-shot, mirroring explorative evaluation where real preferences are yet unknown. Few-shot prompting or fine-tuning, while capable of aligning outputs with desired outcomes, can be seen as priming that does not guarantee generalization to different people or tasks to reflect genuine preferences \citep{gao2025, brand2024}.

Iterative refinement with 14 studies ensured that our prompts achieved more consistent results, given LLMs’ struggle with following instructions and brittleness against minute prompt changes, especially in prolonged conversations \citep{kovac2024, gao2025}. Final prompts repeated directives relevant across multiple steps, such as the format of selecting designs in the preference task as shown in Appendix A. Responses from mega-personas were generated in batches by 20 and then merged. This was not motivated by output size (32K for GPT 4.1, 128K for GPT 5.2), but rather increased errors in long outputs (e.g., missing answers, wrong format). To optimize execution time, past image inputs were removed from the context when irrelevant.

\subsection{Measures}

To assess LLMs’ capacity to predict human design preferences, and the prospect of simulation parameters to positively moderate the relationship between synthetic and real responses, we used a comprehensive set of comparative measures. Our measures, shown in \autoref{tab:measures}, were calculated at the task level to capture various aspects of multinomial preference choices as well as their open-ended text justifications. LLMs can project high surface-level complexity while lacking personally authentic narrative depth \citep{venkit2025}. Therefore, to delve past linguistic elaboration that could disguise simple and generic narratives, we performed a deep qualitative analysis of the semantics and variability of justifications.

\begin{table}[!ht]
\caption{Comparative measures.}\label{tab:measures}
\begin{tabular*}{\textwidth}{lp{12cm}}
\toprule
\textbf{Measure} & \textbf{Description} \\
\midrule
\multicolumn{2}{l}{\textbf{Preference Similarity (choice task)}} \\
\midrule
\begin{tabular}[t]{@{}l@{}}\emph{Distribution Difference}\\ Values: true/false\end{tabular} & Chi-squared test test of goodness of fit applied to determine statistical difference in preference frequencies. \\
\begin{tabular}[t]{@{}l@{}}\emph{First-choice Agreement}\\ Values: true/false\end{tabular} & Match between the most popular preferences. Practical measure to indicate the accuracy of the best option across tasks. \\
\begin{tabular}[t]{@{}l@{}}\emph{Rank-order Agreement}\\ Values: 0-1\end{tabular} & Proportion of response options that appear in matching rank order in resulting preferences. \\
\begin{tabular}[t]{@{}l@{}}\emph{Jensen-Shannon Distance}\\ Values: 0-1\end{tabular} & Quantified differences between probability distributions to capture shifts and enable implementation across studies with different sample sizes (unlike Cramér's V). Higher values signify greater differences. \\
\begin{tabular}[t]{@{}l@{}}\emph{Entropy}\\ Values: 0-1\end{tabular} & Normalized entropy, reflecting the variability of preferences. Higher values signify more uniform distributions, lower values indicate concentration in just a few options. \\
\begin{tabular}[t]{@{}l@{}}\emph{Unique Selection Count}\\ Values: Integer\end{tabular} & The number of responses selected by at least one participant. Complementary with entropy to facilitate interpretation. \\
\midrule
\multicolumn{2}{l}{\textbf{Justification Similarity (open-text)}} \\
\midrule
\begin{tabular}[t]{@{}l@{}}\emph{Lexical Similarity}\\ Values: 0-1\end{tabular} & Cosine similarity between centroid vectors of human and synthetic responses in a shared Term Frequency-Inverse Document Frequency (TF-IDF) space \citep{lumintu2023}. \\
\begin{tabular}[t]{@{}l@{}}\emph{Semantic Similarity}\\ Values: 0-1\end{tabular} & Cosine similarity between the centroids of human and synthetic sentence embeddings generated with the pretrained all-MiniLM-L6-v2 transformer model. \\
\begin{tabular}[t]{@{}l@{}}\emph{Semantic Diversity}\\ Values: 0-1\end{tabular} & Average pairwise cosine distance between pretrained all-MiniLM-L6-v2 transformer model embeddings of all open-text responses \citep{venkit2025}. Higher values indicate broader thematic variation and less redundancy. \\
\begin{tabular}[t]{@{}l@{}}\emph{Lexical Repetitiveness}\\ Values: Real number\end{tabular} & Yule’s K captures repetitiveness of the vocabulary, with higher values demonstrative of lower lexical diversity \citep{hohne2024}. \\
\begin{tabular}[t]{@{}l@{}}\emph{Readability}\\ Values: Real number\end{tabular} & Flesch reading ease estimates how challenging it is to read the text \citep{hohne2024} as an indicator of comprehensibility. \\
\begin{tabular}[t]{@{}l@{}}\emph{Word Count}\\ Values: Integer\end{tabular} & Number of words in an open-text response.\\
\bottomrule
\end{tabular*}%
\end{table}

\section{Results}

\subsection{Synthetic preference evaluation (H1)}
\label{sec:h1}

We compared synthetic and real preferences in all N = 78 tasks. Chi-squared tests return different distributions in 44\% tasks, with 53\% agreement on the first choice. Average rank order agreement was 0.53 (\textit{SD} = 0.46). Jensen-Shannon distance offers a more nuanced perspective, indicating minor to moderate gaps between patterns depending on the task (\textit{M} = 0.17, \textit{SD} = 0.11). Normalized entropy was higher for synthetic preferences (\textit{M} = 0.93, \textit{SD} = 0.07) than for real preferences (\textit{M} = 0.86, \textit{SD} = 0.15), $z = 3.68, p < .001, r = .29$, demonstrating that LLMs inflate indecisiveness when humans have clearer favorites. This is not universal, as in some rarer cases, humans were evenly divided while LLM choices were variation-free. These divergences between synthetic preferences and the ground truth support \hyperref[hyp:h1]{H1}.

\subsection{Moderation by model selection (H2a)}

Using the newer model GPT 5.2 yields fewer significant distribution differences 38\%, although the difference from GPT 4.1 is not significant, $\chi^2(1, 156) = 0.24, p = .63$. First-choice agreement mirrors this at 65\%. Other measures also show a lack of significant change when using the reasoning model at $p > .05$. The data thus corroborates the predicted failure of GPT 5.2 to improve the fidelity of reflected preferences (\hyperref[hyp:h2a]{H2a}).

\subsection{Moderation by sampling parameters (H2b, H2c)}

Neither temperature nor nucleus (top\_p) sampling parameters had significant impact on generated preferences. For temperature = 0.2, distribution difference from real preferences was significant in 41\% tasks, $\chi^2(1,156) = 0.03, p = .87$. For top\_p = 0.2, it was comparably unchanged at 38\%, $\chi^2(1,156) = 0.24, p = .63$. Notably, the issue detected with entropy — LLMs’ lack a firm preference, exaggerating the plausible validity of less popular choices — was not improved by reducing randomness (temperature: \textit{M} = 0.94, \textit{SD} = 0.06; top\_p: \textit{M} = 0.94, \textit{SD} = 0.06). Therefore, \hyperref[hyp:h2b]{H2b} and \hyperref[hyp:h2c]{H2c} are supported, as temperature and top\_p have a minimal effect on simulated preferences.

\subsection{Moderation by persona type (H2d)}

The output of the single-persona approach diverged from real responses significantly more than in the mega-persona variant. Chi-squared was significant in the overwhelming majority of tasks (91\%), making their differences significant compared to mega-personas $\chi^2(1, 156) = 37.75, p < .001, V = .49$. Jensen-Shannon distance reflects this as a gap that was significantly greater than for mega-personas \textit{M} = 0.45, \textit{SD} = 0.16 ($z = -9.14, p < .001, r = .73$). An explanation for this is offered by significant reduction in entropy, \textit{M} = 0.28, \textit{SD} = 0.31 ($z = 9.95, p < .001, r = .8$), which demonstrates more deterministic and less varied responses. This is further reflected by the restrictiveness of the unique selection count, \textit{M} = 1.82, \textit{SD} = 0.83 (mega-persona for comparison: \textit{M} = 2.59, \textit{SD} = 0.83). Although persona type moderates preferences, the better performance of mega-personas is only relative to single-personas, as was illustrated by the analysis in section \ref{sec:h1}. As such, the evidence supports \hyperref[hyp:h2d]{H2d}.

\subsection{Moderation by persona specificity (H2e)}

Simulation of preferences with generic featureless personas resulted in remarkably similar preferences to complex specific personas modeling demographics, personalities and other available information about the original samples. Distributions were different in 46\% of tasks, which is not significantly different from the baseline, $\chi^2(1, 156) = 0.03, p = .87$. Almost identical Jensen-Shannon distance and entropy corroborate a comparably flawed fit. \hyperref[hyp:h2e]{H2e} is supported as the introduction of detailed persona does not systematically improve the generated preferences.

\subsection{Synthetic justification evaluation (H3)}
\label{sec:h3}

To gain insights about factors LLMs associate with their preferences — and how these relate to human reasoning — we conducted a statistical analysis of both open-ended and close-ended justifications, followed by a qualitative examination. The complete evaluated measures, Mann-Whitney and chi-squared test results are presented in Appendix B.

\begin{figure}[ht]
\centering
\includegraphics[width=\textwidth]{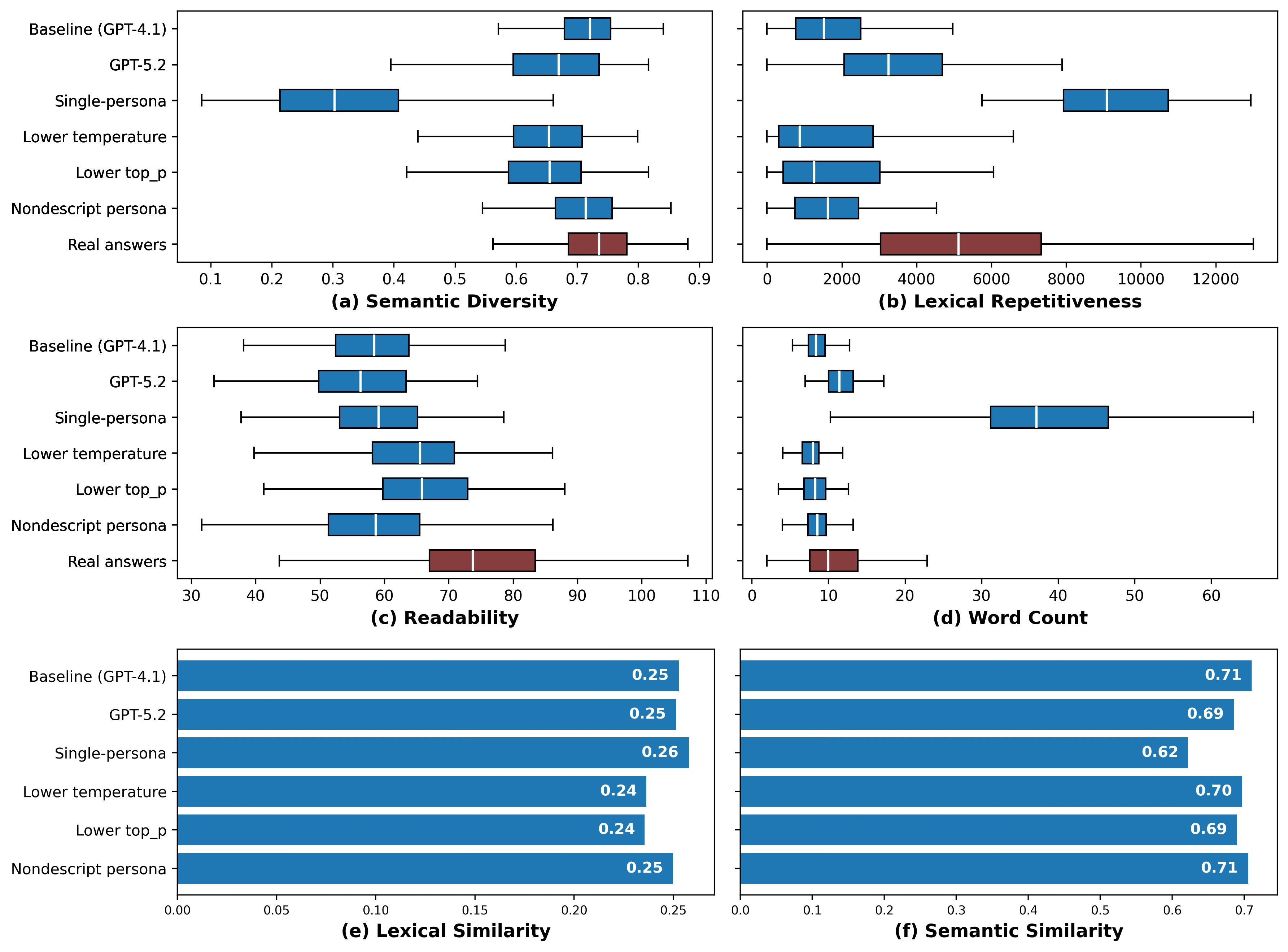}
\caption{Open-ended justification measures (a-d) and linguistic similarity of simulations to real justifications (e, f).}\label{fig:open-justif}
\end{figure}

Key comparisons between real and synthetic open-ended justifications are visualized in \autoref{fig:open-justif}. Significant word count differences could be mechanistically tied to batch sizes (20 in mega-persona approaches, 1 in single-persona), as GPT generated significantly longer answers when simulating individuals, although the outputs did not approach the max token limit in either variant. The single-persona answers were the most different from real answers overall, being lexically repetitive but lacking semantic diversity. Mega-personas, especially in the baseline setting or with nondescript personas, achieved semantic diversity comparable to humans, but significantly smaller readability and lexical repetitiveness.

Further examination of average lexical and semantic similarity to real responses ($\sim$0.25 and $\sim$0.70 respectively), shown in \autoref{fig:open-justif}, indicates small overlap between vocabularies, accompanied by moderately intersecting semantics that contain meaningful differences. For reference, against the baseline, the lexical and semantic similarities of the variants were $\sim$0.4 and $\sim$0.9 respectively. Detailed thematic analysis identified that justifications diverged from humans regardless of whether preferences in the related task matched or not.

Through qualitative analysis of justifications, we uncovered several recurring patterns explaining these divergences. Aside from fundamental differences in frequent words and themes, many LLM feedbacks and suggestions were unequivocally nonsensical. Although LLMs could competently extract and describe visual elements in stimuli, they failed at common-sense interpretation and contextual awareness. Examples include praising the usefulness of purely illustrative elements for a task the models hallucinated, or derailment toward aspects irrelevant for the design. 

Genericness was another common issue. Although human answers were not bereft of general statements, they were often supported by focus on specific differences between stimuli (e.g., why they prefer placement of an element on left and right), LLMs often rationalized their choice with general concepts such as clarity or “vibe”. GPT responses often took the form of a generic positive followed by a generic negative without a clear logical link, such as “Visibility is great, but makes me hesitant to recommend”. Other times they picked a specific element and made a generic comment about it, such as praising the visibility of buttons that were the same across all options. Human responses rarely focused on just one specific visual element, bur rather key utilitarian or hedonic value of information inferred from the stimulus as a whole.

Other distortions reflected LLMs’ lack of human experience and subjectivity. Humans often brought up their emotions, feelings and perceptions, such as their views of objects as personal, human or natural. LLMs sometimes faked vague prior experiences, such as expressing familiarity with the design “from other websites”, but with no clarification. Unlike humans, LLMs did not complain about unfamiliar or less intuitive word choice but took them at face value. Various human behaviors did not manifest in LLMs, as they run counter to their bias to helpfulness, harmlessness and honesty. They did not voice open dislike to all available options, nor discuss personal dislikes and aversions (e.g., not wanting to share information noticed in a design). Instead, many LLM justifications resembled impersonal guidelines (e.g., general suggestion to improve accessibility to non-native speakers).

Stimulus complexity and nuance broadened the gap between real and synthetic preferences. In visually complex stimuli (e.g., dashboards), LLMs hyperfocused on specific random elements as the fulcrum of their argument, even as they were less likely to be relevant. In copywriting alternatives communicating the same message differently, LLMs failed to capture nuanced subjective reasons that caused some options to resonate with people more strongly

Inconsistency with human justifications also translated into close-ended answers. Distributional difference was found in 53\% of answers, with Jensen-Shannon distance of \textit{M} = 0.21, \textit{SD} = 0.11 and entropy \textit{M} = 0.84, \textit{SD} = .14. Between model variants, the only significant difference was in order agreement, which was slightly higher for GPT 4.1 (\textit{M} = 0.48, \textit{SD} = 0.33) than 5.2 (\textit{M} = 0.35, \textit{SD} = 0.35), $z = 2.04, p = .039, r=.21$. Given the scope and pervasiveness of the misalignments discussed above, \hyperref[hyp:h3]{H3} is supported.

\section{Discussion}

In this research, we identified design risks that emerge for society and organizations with the increasingly ubiquitous use of large language models by practitioners online \citep{hui2024, niedehoffer2025}. Training on human data like reviews might make LLMs tempting as a source of rapid feedback. However, misunderstood or disguised as a human-centered approach, they can compromise designers’ ability to comprehend and address actual design preferences even in simple scenarios. To examine whether and how LLMs can mislead decisions by organizations, designers, product owners and other stakeholders, we examined 29 preference tests conducted by different organizations in real practice. In our systematic study, we compared real preference data to LLM simulations. 

Our findings demonstrate that LLMs fail to predict design preferences. Their rationalizations of the “why” range from misaligned through generic to nonsensical. Plausibility, elaborativeness and limited similarity mask an inability to account for deep and nuanced factors that influence real preferences. Our evaluation of reactions to visual designs aligns with and expands upon previous studies that showed LLMs struggle to represent people’s moral, economical, consumer and political preferences \citep{durmus2024, qu2024, santurkar2024, park2024c, brand2024, yoon2024}. Through comprehensive assessment of moderating variables that encompass model selection, configuration, and prompt engineering, we show that the issues are not solved by implementing detailed personas or chain-of-thought mechanisms in Large Reasoning Models like GPT 5.2.

\subsection{Theoretical implications}

While prior research investigated misalignment between LLMs and preferences in broader social contexts or toward text descriptions of products \citep{durmus2024, qu2024, santurkar2024, park2024c, brand2024, yoon2024}, our study advances the knowledge by focusing on simulated formation of preferences toward visual design stimuli. The evidence for our hypotheses confirms that LLMs informed about the conditions of concrete sensory experiences cannot account for latent internal and external factors meaningfully affecting human samples (e.g., perception of design content, attitudes, conventions). Low lexical repetitiveness and readability are demonstrative of a style that is eloquent and elaborate, yet lacks narrative depth \citep{venkit2025}. We identify two types of misleading outcomes in synthetic preferences: incorrect order, and exaggerated balance. Lenient preferences — also encountered by \citet{yoon2024} — could be attributed to the homogenizing effect and positive bias \citep{lazik2025}.

Our findings advance the theoretical understanding of relevant dimensions for measuring algorithmic fidelity in LLM-generated responses. \citet{venkit2025} employed creativity measures like semantic diversity while highlighting the importance of narrative depth and variability over elaborative complexity. Although semantic diversity can indicate significant gaps in generated personas \citep{venkit2025}, we find no such difference in preference justifications. We argue that this stems from the greater depth of self-descriptions, which encompass complex life experiences, views, skills, etc. For design stimuli, comparable semantic diversity can be caused by three factors: a) the inherently smaller answer space, b) exaggerated preference entropy that increases diversity even when answers remain generic and c) hallucinated nonsensical or misaligned answers. Within constrained topics, diversity measurement should not be subject to content-agnostic interpretation, but studied with awareness of nuance, depth, and fidelity. Qualitative analysis retains a key role for understanding contextual, subtextual, and logical aspects of intents and ideas.

By analyzing justifications, we have further corroborated research exposing reasoning gaps between humans and LLMs. This manifests as the inability of LLMs to extract associations and interpret sensory stimuli \citep{imschloss2025}, along with their struggle with causal reasoning \citep{shojaee2025}. According to reason-based theory of rational choice, preferences are driven by individual and context-sensitive reasons \citep{dietrich2013}. This is evident from preference justifications. Although some human responses can be sparse, others clearly demonstrate inference of nuanced and diverse high-level patterns, supported by logical narratives and resulting in judgments. Answers can combine such information, keep it implicit, or link to other sentences through text and subtext (e.g., “Option is easier to understand because its layout is symmetrical and not overly crowded.”). LLMs supplant reasoning for stochastic elaboration of linguistic associations, hence their lapses in common sense, flattened rationality and jumping to generic conclusions (see \ref{sec:h3} for examples).

Our exploration of moderating variables demonstrates that wider problematic patterns generalize across configurations, including solutions aimed at improving model reasoning and representativeness. The introduction of iterative chain-of-thought in GPT 5.2 does not confer notably better alignment. Through attempts to correct outputs, it even compounds the model’s biases, as manifested through exacerbated deviations and global reduction in semantic diversity in justifications. The persistence of reasoning gaps strengthens the argument regarding the illusory nature of LLM “reasoning” \citep{shojaee2025}. Specific personas describing the sample provide no tangible benefit or detriment. With LLMs’ leaning on stereotypes and struggling to simulate subgroups \citep{park2024c, brand2024}, this points to an inability to realistically capture perspectives of different people alongside the impacts of their latent characteristics. It is credible that the better alignment when simulating populations (mega-personas) arises from the encoded expectation for greater diversity rather than a fundamentally richer and context-aware understanding of people.

\subsection{Implications for practice}

Our research carries key implications for practitioners and organizations already using or looking to incorporate LLMs into their processes to generate design feedback, supplement research, and support decision-making.

As a general guidance for user-centered design and other areas concerned with human reactions to stimuli, LLMs should be avoided as a source of attitudinal feedback. Across a variety of domains and designs, our study showed significant deviation of simulated preferences and their reasons from human outcomes. At various levels of decision-making, machine biases may foreground unpreferrable options. By creating the illusion of balance where all choices are equally good, they can also entrench practitioners in their personal beliefs and biases, discouraging interest in data. Generic grasp of preferences may steer designs to please a fictional general user, but fail to capture the interest and provide satisfaction to real audiences. Due to the severe risks that AI overreliance poses by stunting behavior, cooperation and cognitive abilities \citep{klingbeil2024}, organizational AI policies should actively educate personnel about these limitations. Approaches adopted to study preferences should verifiably foster or augment empathy.

As shown through systematic investigation, various parameter configurations, specific personas or iterative chain-of-thought in GPT 5.2 fail to achieve significant breakthroughs. For seekers of productive AI-driven practices, these findings invite skepticism toward the conception that LLMs can self-reliably provide responses reflective of people described to them via prompts, instead of stochastically parroting their training data \citep{argyle2023, gao2025}, or that new LLMs will become more human-like without a significant paradigm shift.

Lastly, in spite of its flaws, the believability of LLM-generated design preference data is sufficient to raise the concern about fake data among research experts and online research platforms \citep{sop2024}. Real datasets could be polluted by bots and malicious actors, motivating the development of measures for mitigation in justifications (in addition to standard text-based methods). Our research makes a significant contribution by identifying high-level properties that could help indicate LLM fakeries: disregard for key differences between stimuli; hyperfocus on one or multiple specific visual elements; assignment of generic properties to the stimulus or visual elements without discussing implications; elaboration without adding depth or nuance; lack of narrative between statements; overpraising or precisely balanced praise and criticism; lack of subjective evaluations; assumptions incongruous within the context; off-topic suggestions; and inconsistencies.

\subsection{Limitations and future work}

Our study has limitations related to its methodological context. This includes the innate biases of preference testing as a self-reported method where the presence of multiple options may have confounding effects. Since AI-based solutions can struggle with adapting to complex and diverse conditions and tasks \citep{kuric2025et, kuric2025lies}, ecological validity was the chief priority for our method. Our collected data originates from real practice of various organizations, with studies that were highly diverse in the phrasing of tasks and a heterogeneity of justification questions. This was suitable for our study’s goals to produce generalizable analysis of real-world patterns. Systematic biases due to uncontrolled variables in study protocols were mitigated by random selection of a reasonably large sample of studies. Future research could use more controlled conditions to investigate the effects of study context.

Access to the study data was provided by data owners exclusively for the purpose of this research. Since the studies constitute sensitive proprietary data protected by privacy, results are presented in aggregated form where no information about data owners or specific contents of their studies can be inferred. Although high external and ecological validity of data is guaranteed, it sets a limit on reproducibility. For a balanced approach to transparency and data protection, full statistical analyses of preferences and justifications (open- and close-ended) are provided in Appendix B.

\subsection*{Acknowledgements}
This work was supported by the EU NextGenerationEU through the Recovery and Resilience Plan for Slovakia under the project No. 17I04-04-V05-00029, and co-financed by the Slovak Research and Development Agency under Contract No. APVV-23-0408. We would like to thank the UXtweak Research team for their technical and expert support.

\subsection*{Ethical Statement}
All procedures were performed in compliance with relevant laws and institutional guidelines and have been approved by the appropriate institutional committee. The privacy rights of human subjects have been observed and that informed consent was obtained for experimentation with human subjects.

\subsection*{Declaration of competing interests}
The authors have no relevant financial or non-financial interests to disclose.

\setlength{\bibsep}{0pt}
\bibliographystyle{apacite}
\bibliography{sources}

\end{document}